\documentstyle[aps,epsfig]{revtex}

\title{Normal state pseudogap and $(\pi,0)$ feature
in the underdoped 
  high-$T_c$ cuprates. A microscopical theory.}

\author{F Onufrieva and P Pfeuty}
\address{Laboratoire Leon Brillouin CE-Saclay 91191 Gif-sur-Yvette France}

\begin{document}

\twocolumn[\hsize\textwidth\columnwidth\hsize\csname@twocolumnfalse\endcsname
\maketitle

\begin{abstract}
We show that a simple 2D electron system on a square
lattice with hoping between more than nearest neighbors exhibits
in the presence of electron spin exchange interaction 
properties strikingly similar to those observed in the 
underdoped cuprates It is a normal state pseudogap, its behaviour
 with doping and $T$, a form of the spectrum around
$(0,\pi)$ etc. The motor is the electronic topological
transition  which due to two-dimensionality gets
new features with respect to well-studied 3D case. 
\end{abstract}
]

\date{\today}

Many experiments  for the hole doped high $T_c$ cuprates
provide an evidence for the existence of a pseudogap
in the
underdoped regime above $T_c$, see for example
 \cite{NMR,conductivity}.
 Its value  increases
 with increasing
 doping distance
from the optimal doping,
$\delta_{opt}-\delta$, the pseudogap exists
until rather high temperature
$T^*_{gap}(\delta)$
which increases in the same way with $\delta_{opt}-\delta$
 while the critical temperature
of superconducting (SC) transition decreases
that makes  doubtful its SC origin.
 ARPES experiments \cite{Marshall,ARPES,Maekawa} show  that  the gap opens
 for wavevectors
around $(0,\pi)$,  the spectrum gets a flat 
shape and it "stops" at some threshold wavevector
 in the direction
$(0,\pi)$-$(\pi,\pi)$, the spectral functions 
have very damped form for $\bf k$ in the vicinity
of  $(0,\pi)$ (so called
 $(0,\pi)$-features).
The problem of the electronic anomalies  above $T_c$ 
is considered today as a
key problem which should help to better understand the
microscopic origin of  high-$T_c$ superconductivity.

Two approaches are possible to attack the problem.
A phenomenological one which uses an experimental information,
 for example that information 
(from neutron scattering and NMR) which tells us that significant
bosonic field (corresponding to magnetic fluctuations)
exists above $T_c$ just in the underdoped regime. This
pushes up the idea that the anomalous electron spectrum can be
a result of fermion-boson interaction. The
theories \cite{Schriffer,Pines,Chubukov} are of such type; the  bosonic
fluctuations are presumed to be a memory
about AF insulating state existing at low doping. There are
also theories  which consider bosonic field related to
SC instability \cite{Levin} and polaron effect \cite{Ranninger}.

The second type of approach, microscopical, should {\bf obtain} the
bosonic field if it is a fermion-boson interaction
which is at the origin of the phenomena
and should explain {\bf why this bosonic field is strong in
the so wide range of doping and temperature} quite far from both 
 AF
and SC ordered phases.
 This way is of course preferable as 
it allows to reveal a nature of effects. The theory
which we start to develop in \cite{OnPfeuty} is of the second type.
We have considered a  metallic
 state in 2D electron system on a square lattice. We have shown firstly that
an electronic topological transition (ETT)
 occuring in the system of noninteracting
electrons  at  electron
concentration $n_c$ when Fermi level (FL) crosses 
saddle-point (SP) is quite untrivial
in the case of   hoping between more than nearest neighbours,
$t' \neq 0$ or/and $t'' \neq 0$,... (more general in
all cases excluding the perfect nesting).
Corresponding quantum critical point (QCP) $\delta=\delta_c$, $T=0$
($\delta=1-n$)  combines
two independent aspects  of criticality, the ordinary one
related to
 logarithmic singularities
in thermodynamical properties and the untrivial one:
this point is the end point of the critical line
related to Kohn singularities in 2D electron polarizability
\cite{OnPfeuty}.
Quite trivial consequence of
this ETT is a developing of density wave (DW) and SC instabilities around
the point $\delta=\delta_c$, $T=0$ in the presence of interaction
and therefore an existence
of critical bosonic fluctuations in the disordered
metallic state above them. The untrivial aspects
concerning to DW degrees of freedom  are:
(i) strong assymetry between regimes $\delta<\delta_c$ and
$\delta>\delta_c$ and (ii)
  very long (in doping and temperature) memory about
DW instability in the disordered state on one side of ETT,
$\delta<\delta_c$.  This encourrages us to
continue our analysis and to consider now the interaction
of bare electrons with DW bosonic field
that is the aim of the present paper.

The basic equations are following. The spectral function
corresponding to the renormalized Green function,
$A({\bf k},\Omega)=\lim_{\gamma \rightarrow 0}|Im G(({\bf k},\Omega
+i\gamma)|$, is given by

\begin{equation}
A({\bf k},\Omega)=\frac{|Im\Sigma({\bf k},\Omega)|}
{[\Omega-\epsilon_{{\bf k}\sigma} - Re\Sigma({\bf k},\Omega)]^2
+[Im\Sigma({\bf k},\Omega)]^2}
\label{1}
\end{equation}
with the irreducible part $\Sigma
({\bf k},i\Omega_n)$  determined as
$$
\Sigma
({\bf k},i\Omega_n)=
\sum_
{\bf p}
\int
\frac{d\omega}{\pi} (J^2_{\bf p}) 
Im\chi({\bf p},\omega)
[\frac{
 n^B(\omega)
+n^F(\tilde\epsilon_{{\bf p+k},\sigma})}
    {i\Omega_n+\omega -
       \tilde\epsilon_{{\bf p+k},\sigma}}$$
\begin{eqnarray}
+\frac{1+ n^{B}(\omega)-n^{F}(\tilde{\epsilon}_{{\bf p+k},\sigma})}
{i\Omega_n-\omega -\tilde{\epsilon}_{{\bf p+k},\sigma}}].
\label{2}
\end{eqnarray}
[We work with electron spin exchange  interaction,
the same  which leads to SC instability \cite{Onufrieva}  and
SDW instabilty \cite{OnPfeuty}]
 Being practically standard for the case of fermions
interacting with bosonic field, the equations lead to results which explicit
form crucially depends on   explicit forms of the bare electron
 spectrum $\epsilon_{{\bf k}\sigma}$ and
 of the bosonic Green function (or corresponding susceptibility
$\chi$).
In our  approach we use
 the susceptibility {\bf calculated} for the considered 2D electron
system  in the disordered state above both SDW and SC
 instabilities. And  we use the spectrum
which posseses SP : in the form

\begin{equation} 
\epsilon_{{\bf k}\sigma} =
\ -2t (cos k_x + cos k_y) - 4t' cos k_x cos k_y
\label{4}
\end{equation}
for numerical calculations and in the  form

\begin{equation}
\tilde\epsilon_{{\bf k}\sigma} = \epsilon_{{\bf k}\sigma} -\mu
= -Z +a k_x^2 -
b k_y^2,
\label{3}
\end{equation}
for analytical estimations. In (\ref{3}), 
$k_x$, $k_y$ are distances from SP wavevector
$(0,\pi)$,
$a=t-2t'$, $b=t+2t'$.
 The important parameter is
 $Z$,  
defined as $ Z=\mu-\epsilon_{s}$
($\epsilon_{s}=-4t'/t$ is SP energy) 
and measured
the energy distance from ETT : $Z \propto \delta_c-\delta$ \cite{OnPfeuty}.
This parameter determines a {\bf new energy scale in the system}.
 As we will see, 
{\it results} for $\Sigma$ and therefore
for the renormalized electron Green function
{\it are very different from those obtained in \cite{Chubukov,Pines}
where the SP in the bare spectrum was ignored}.

As shown in \cite{OnPfeuty}, the susceptibility in the regime $\delta<\delta
_{DW}(T=0)$ of the disordered metallic state
can be roughly presented in the form

\begin{equation}
\chi({\bf q},\omega)=\frac{1}{4J}
(\frac{(1-\kappa^2)+iC(\omega) \omega}{\kappa^2+
A({\bf q-Q}_{AF})^2-iC(\omega) \omega})\label{5}
\end{equation}
valid  for low energies $\omega \ll \omega_c \propto \delta_c-\delta$,
with $C(0)\propto 1/Z$,
the parameter
$\kappa^2$ describes a proximity
to the  DW instability line.
The important parameter is the energy
$\omega_0=\frac{\kappa^2}{C(0)}$ for which $Im\chi({\bf q},\omega)$
is
 maximum, the parameter  which determines {\bf the 
second energy scale in the system}.
  The behaviour of 
 $\kappa^2$,  $C(0)$, and  
 $\omega_0$, as functions of $\delta$ and $T$
is quite untrivial, we discuss this point later on.
We use the form (\ref{5}) for analytical estimations while RPA form
with the electron-hole loop
calculated based on the full spectrum (\ref{4})  
\cite{OnPfeuty}  for numerical calculations.

\begin{figure}
\epsfig{%
file=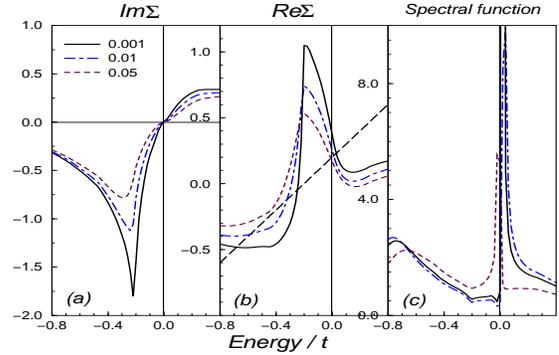,%
figure=fig1.eps,%
height=4.5cm,%
width=7.5cm,%
angle=0,%
}
\\
\caption{$Im\Sigma$,  $Re\Sigma$ and spectral functions 
as  functions of $\Omega$ for ${\bf k}={\bf k}_{SP}$
 and different values of $\kappa^2=0.05, 0.01, 0.001$
depending on $t/J$.
 ($T=0$,  $Z/t=0.2$, $t'/t=-0.3$.) }
\label{f1}
\end{figure}

Analytical calculations for $T$=$0$ performed
with the  spectrum (\ref{3})
and susceptibility (\ref{5})
 for $\bf k$=${\bf k}_{SP}$=$(0,\pi)$ give for
 $\Omega$ corresponding to the bare spectrum  
$\epsilon_{{\bf k}_{SP}}=-Z$

\begin{eqnarray}
Im\Sigma
({\bf k}_{SP},\Omega=\epsilon_{{\bf k}_{SP}}) \propto
 -\frac{1}{ C(0) }
\ln(\frac{ZC(0)}{\kappa^2}) \ln(\frac{2C(0)}{m})
\label{21}
\end{eqnarray}
 ($m=(1/a+1/b)/2$) while for 
low energies
$\Omega \ll \omega_0$  they give : $|Im\Sigma
({\bf k},\Omega)| \propto \Omega^2$.
Numerical calculations performed with the full spectrum (\ref{4}) and 
with the calculated RPA susceptibility
give the results shown in Fig.\ref{f1}a.
Both numerical and analytical calculations show that
$Im\Sigma$ taken for fixed $Z$ becomes singular at $\Omega =
\epsilon_{\bf k_{SP}}=-Z$
when $\kappa^2 \rightarrow 0$ and that for very low energies
 the Fermi liquid behaviour survives.
The corresponding behaviour of $Re\Sigma$ is presented in Fig.\ref{f1}b
together with the line $\Omega-\epsilon_{\bf k}$ since
the equation
$\Omega-\epsilon_{\bf k}-Re\Sigma({\bf k},\Omega)=0$
 determines  new poles.
One can see  that there are three poles for small
$\kappa^2$.
The position of the pole at the intermediate energy
 is close to the position of
the pole in the bare spectrum while the two external poles are 
precursors of the new bands in the ordered DW
phase, see \cite{Kisselev}.
The important point is that  the spectral
function   
 exhibits {\bf maxima} approximately at the energies corresponding to
the {\bf new  poles} while it is {\bf minimum at the position
of the bare pole}, see  Fig.\ref{f1}c. This occurs due to the singularity
 (or maximum in the case of finite $\kappa^2$) in $|Im\Sigma|$ 
 just at $\Omega$ corresponding to the position
of the bare spectrum.
  Such a behaviour means an appearence of two new modes
of the spectrum instead of one in the bare spectrum.
For intermediate $\kappa^2$
the poles at negative $\Omega$ disappear, however the
peak in the spectral function at negative $\Omega$
survives being now a resonance peak.

\begin{figure}
\epsfig{%
file=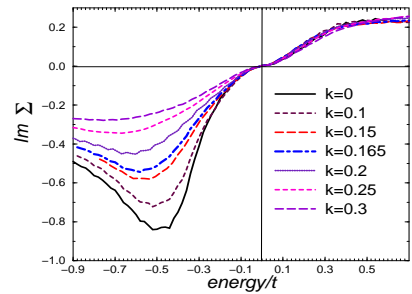,%
figure=fig2.eps,%
height=4.5cm,%
width=6cm,%
angle=0,%
}
\caption{$Im\Sigma({\bf k},\Omega)$ 
as functions of $\Omega$ for 
 wavevectors in the direction $\pi(1-k,k)$.
$T$=$0$, $Z/t$=$0.3$, $t/J$=$1.8$ ($\kappa^2$=$0.066$, $\omega_0/t$=
$0.08$}
\label{f2}
\end{figure}

Calculations of $Im\Sigma$ 
for $\bf k$ 
 along different directions $\phi$ traversing SP show that
 the effect of the maximum in $|Im\Sigma|$
is preserved for wavevectors located in some distance $\Delta k(\phi)$
 from $(\pi,0)$,
see for example Fig.\ref{f2} corresponding to the direction
$\pi(1-k,k)$ which traverses the "hot spot".
The $\Omega^2$ behaviour at low $\Omega$ is also preserved.
[One should note that there is no  square-root
behaviour \cite{Chubukov} for $\Omega>\omega_0$
in the presence of SP.]
This picture is valid for all directions including $(0,\pi)$ - $(\pi,\pi)$
and
$(0,\pi)$ - $(0,0)$. The maximum of $|Im\Sigma|$ (when exists)
occurs at $\Omega^*=-[Z+a\omega_0+b(\Delta k(\phi))^2]$. Above
some threshold value of $\Delta k^*(\phi)$, $Im\Sigma$ becomes 
almost unstructured.

\begin{figure}
\begin{center}
\epsfig{%
file=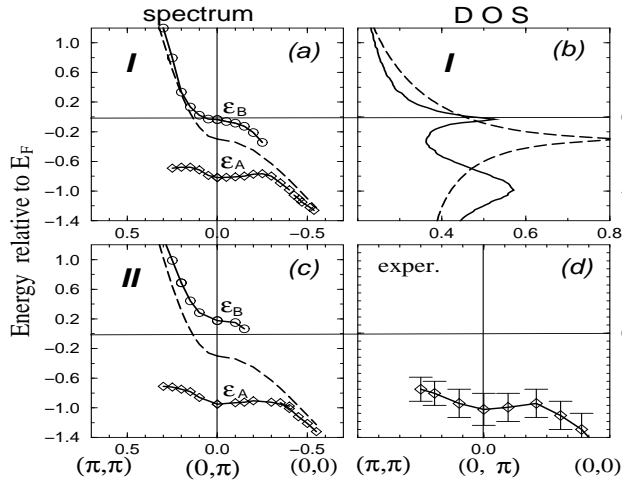,%
figure=fig3.eps,%
height=6.cm,%
width=9. cm,%
angle=0,%
}
\end{center}
\caption{Electron spectrum  
along directions $(0,\pi)-(\pi,\pi)$ and $(0,\pi)-(0,0)$
 in two temperature
regimes, I  ($T$$\ll$$\omega_0$,  $\omega_0$$\ll$$1$)
and II  ($T>\omega_0$)   for fixed $Z$.
Calculations are
done for  (a) $T$=$0$  and (c) $T/t$=$0.16$, in both cases
for $Z/t$=$0.3$ 
($\delta=0.1$).
  For the regime I the corresponding 
DOS 
is shown for comparison (b). Results for the regime II
are compared with ARPES data [3] for  the underdoped BSCO
(d).  The dashed lines in (a), (b), (c) correspond to the bare spectrum.}
\label{f3}
\end{figure}

There are number of important consequences of such a behaviour.
The first is the form  of the spectrum in the vicinity of
 $(\pi,0)$ : the bare spectrum splits into two branches,
 and a gap opens, see Fig.\ref{f3}a for 
the spectrum obtained from maxima
of spectral functions. Secondly,  
{\bf  the spliting into two branches} and the 
{\bf gap disappear} above some threshold wavevector 
 {\bf when one goes far away from SP},  the bare spectrum
is restored.  The corresponding density of states (DOS) calculated for
the same conditions as the spectrum in Fig.\ref{f3}a is shown
in Fig.\ref{f3}b. It exhibits a pseudogap and two new peaks instead of
the SP peak in the bare spectrum.

\begin{figure}
\epsfig{%
file=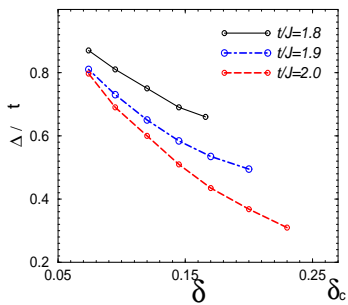,%
figure=fig4.eps,%
height=4.5cm,%
width=5cm,%
angle=0,%
}
\caption{ $T=0$ pseudogap  (determined at $(0,\pi)$) as a function
 of doping for different values of $t/J$ ( $t'/t=-0.3$).
Each curve starts
at $\delta=\delta_{DW}$
 which value is determined
by $t/J$. }
\label{f4}
\end{figure}

Thirdly, being directly related to the energy scale
$Z$$\propto$$\delta_c-\delta$, {\bf the  gap
increases} 
 with $Z$ or by other words {\bf with
decreasing  doping}, see Fig.\ref{f4}. The
coefficient of proportionality depends on interaction
increasing with $J/t$. 
Fourthly, due to   high value of $|Im\Sigma|$
for $\Omega$ corresponding to the lower branch
{\bf this branch  is strongly damped}
in the vicinity of $(0,\pi)$. The damping decreases when one
goes away from SP and the well-defined bare spectrum
is restored. All these features  are in a very
good agreement  with ARPES. There is,
however one important difference:
the upper well-defined branch  characterized
by new SP located   almost on FL  
is not observed
experimentally. 
This could come from finite $T$ effect.

\begin{figure}
\epsfig{%
file=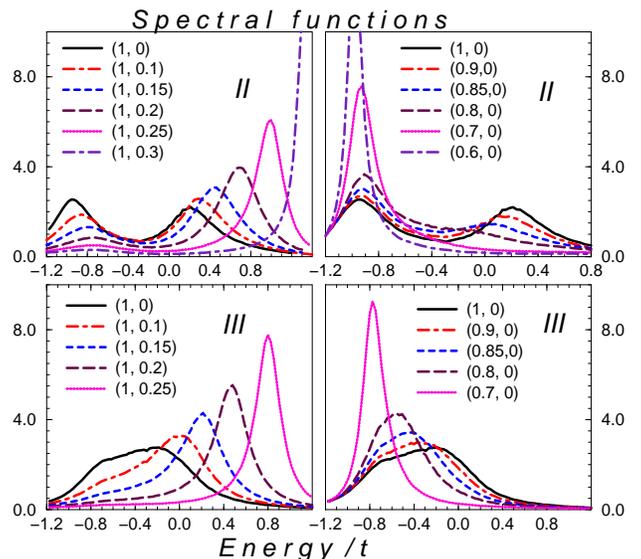,%
figure=fig5.eps,%
height=7cm,%
width=8.3cm,%
angle=0,%
}
\caption{Spectral functions calculated 
for two directions traversing SP
$(\pi,0)$-$(\pi,\pi)$  and $(\pi,0)$-$(0,0)$ 
and for two temperature regimes,  II  and III 
for fixed
doping $\delta$=$0.1$ ($Z/t$=$0.3$). The wavevectors are done in $\pi$ units.
$t'/t$=$-0.3$, $t/J$=$1.8$}
\label{f5}
\end{figure}

To check let's study  a behaviour at  finite 
$T$.
After the analysis similar to that performed  for
$T=0$, one can realize that there are four different
regimes in $T$. For the first three, the lower
branch in the spectrum keeps its main features while the
upper branch changes its behaviour. Namely, in the {\bf low
temperature regime I},
 $T \ll \omega_0$, 
$\kappa^2 \ll 1$, the energy $\epsilon_B$ is negative
and the upper branch traverses FL at some point in the direction
$(0,\pi)-(\pi,\pi)$ as in Fig.\ref{f3}a. 
 {\bf FS exists} being open but quite
 close to the critical form. In the crossover
regime, $ T \sim \omega_0 $, one has
$\epsilon_B \approx 0$, the new SP in the upper branch
traverses FL while this
branch becomes much more damped than in the
regime I. In the regime II, $T > \omega_0$, 
the SP energy moves to positive $\Omega$, $\epsilon_B>0$,
and what is most important the {\bf upper branch does not
traverses FL anymore} (see spectral functions in Fig.\ref{f5}a,
\ref{f5}b
and the spectrum obtained from their maxima in Fig.\ref{f3}c) : {\bf Fermi surface disappears}. And finally in the  regime III,
$T<\omega_0$ while both parameters are not very small,
the spectral functions have a form shown in Fig.\ref{5}c,d. The gap is
filled up, the bare spectrum is restored although
it remains quite damped at $(\pi,0)$.

\vskip 0.2 cm

\begin{figure}
\epsfig{%
file=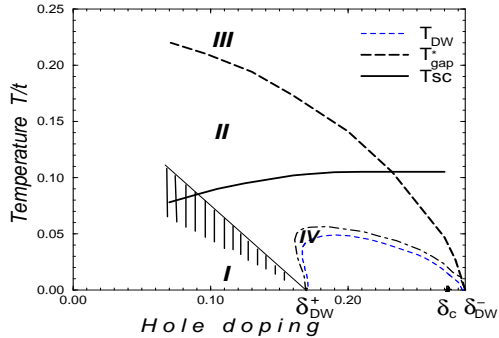,%
figure=fig6.eps,%
height=5cm,%
width=7cm,%
angle=0,%
}
\caption{Calculated phase diagram for $t-t'$ electron
system in the presence of $J$-interaction. $T_{DW}$ is the line of DW
instability [11], $T_{sc}$  of SC instability [12],
different crossover regimes are discussed in the text.
$t'/t=-0.3$, $t/J=1.8$.}
\label{f6}
\end{figure}

Areas in $T-\delta$ plane where each regime takes place depend on
behaviour of $\omega_0$ as a function of $\delta$ and $T$. This
behaviour is analyzed in \cite{OnPfeuty}. We remind that
$\omega_0$ (taken at fixed doping $\delta<\delta_{DW}$),
slightly decreases with increasing $T$ 
below   $T^*_{Re}(\delta)= \alpha 
(\delta_c-\delta)$,  remains constant until another 
temperature, $T^*_{Im}(\delta)= \beta 
(\delta_c-\delta)$ and then increases with $T$. 
This leads to the phase diagram  in
 Fig.\ref{f6}. 
 where  we show one more regime, the regime
IV, $T \gg \omega_0$.
 For it a gap opens for all
$\bf k$ in the direction
$(1-k,k)$,  the spectrum has the form
as in the ordered phase, see Fig.6 in \cite{Kisselev}.
 The origin 
is a peak in $|Im\Sigma({\bf k},\Omega)|$ at $\Omega$=
$\epsilon_{{\bf k+Q}_{AF}}$ existing for all $\bf k$. This
regime
 corresponds to the quasistatic 
regime analysed in \cite{Pines} (although the authors did not note that
the gap opens for all wavevectors along $(k, 1-k)$). It is important to
 emphasize 
that it takes place {\bf only in the intime
vicinity of AF instability} whatever its origin
(SDW or the
spin-localized as in \cite{Pines}).

All regimes discussed have been obtained in ignoring
a superconductivity. In the presence of the latter (SC instability also
develops around the QCP as discussed in \cite{Onufrieva,OnPfeuty})
the regimes I, IV are unacessible being covered by SC phase.
It is the regime II which occupies the most 
interesting part of $T-\delta$ plane  and which should be
compared with  experiment for the pseudogap regime.
The corresponding ARPES
spectrum for the underdoped
BSCO \cite{Marshall} is shown in Fig.\ref{f3}d.
 The agreement is excellent. 

\begin{figure}
\epsfig{%
file=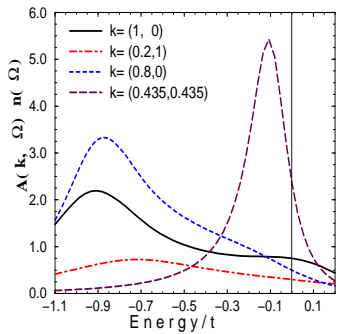,%
figure=fig7.eps,%
height=5cm,%
width=5cm,%
angle=0,%
}
\caption{The function $A({\bf k},\omega) n^F(\omega)$
for $Z/t$=$0.3$, $T/t$=$0.16$ and four characteristic
wavevectors. $t'/t$=$-0.3$, $t/J$=$1.8$}
\label{f7}
\end{figure}

With the same aim
to compare with experiment 
we show in Fig.\ref{f7} the response function
$A({\bf k},\omega)n^F(\omega)$ corresponding to that measured
by ARPES for four most characteristic wavevectors : for
$(\pi,0)$ to emphasize the untrivial
$\Omega$ dependence, for two wavevectors located not far
from $(\pi,0)$ in two directions $(\pi,0) - (\pi,\pi)$ and
$(\pi,0) - (0,0)$ to emphasize  the absence of crossing
of FL,  and for the wavevector close to $k_F$ in $(1,1)$
direction to emphasize the ordinary FL form of the response
function. The behaviour is very close to that observed experimentally
\cite{Marshall}.
All these regimes are related to the DW fluctuations.
 The question arises  about a precursor of the superconductivity.
Do  corresponding fluctuations  succeed in opening  a
pseudogap. The answer is that it is indeed so but only in
the intime vicinity of $T_{sc}(\delta)$ in the same way
as for the regime IV with respect to the line $T_{DW}(\delta)$;
the details will be discussed elsewhere.

Summarizing,
we have found that a simple 2D electron system on a square
lattice exhibits in the presence of electron spin exchange interaction 
properties strikingly similar to those observed in the 
underdoped cuprates, the normal state pseudogap, its behaviour
 with doping, its existence below  $T^*_{gap}$ increasing with
decreasing $\delta$, the shape of the spectrum around
$(0,\pi)$ etc. : the fact
which is quite impressive in a view
that we did not
use any external hypothesis or adjastable parameters.
 The motor is ETT which due to two-dimensionality gets
new features \cite{OnPfeuty} with respect to well-studied 3D case, see
\cite{Lifshitz,ETT}. The
effects disappear in the nesting case, $t'=t''=...=0$: the
regimes $\delta<\delta_c$ and $\delta>\delta_c$ become
symmetrical, for both regimes the bosonic field is incommensurate and
rapidly weakening with increasing $|\delta_c-\delta|$ and $T$ 
\cite{OnPfeuty}. The results are in a good agreement with
exact diagonalization study for $t-t'-J$ model 
\cite{Maekawa}.
The general picture is not
sensitive to values of microscopical
parameters, $t'/t$ and $t/J$ (only positions of the
characteristic points, $\delta_c$,  $\delta_{DW}^+$,
$\delta_{DW}^-$ depend on them). The theory predicts a
quite interesting behaviour for the spectrum and 
spectral functions for positive $\omega$ that would
be interesting to check by photoemission experiment.

\end{document}